\documentclass[aip,rsi,reprint,showpacs,superscriptaddress]{revtex4-1}
\usepackage{hyperref}
\usepackage{amsmath,amssymb,graphicx}
\usepackage{epstopdf}
\usepackage{color}
\usepackage{subcaption}

\begin{document}

\title{Analysis and calibration techniques for superconducting resonators}

\author{Giuseppe Cataldo}\email{Giuseppe.Cataldo@NASA.gov}
\author{Edward J. Wollack}
\author{Emily M. Barrentine}
\author{Ari D. Brown}
\author{S. Harvey Moseley} 
\author{Kongpop U-Yen}
\affiliation{NASA Goddard Space Flight Center, Greenbelt, MD 20771, USA}

\begin{abstract}
A method is proposed and experimentally explored for \textit{in-situ} calibration of complex transmission data for superconducting microwave resonators. This cryogenic calibration method accounts for the instrumental transmission response between the vector network analyzer reference plane and the device calibration plane. Once calibrated, the observed resonator response is analyzed in detail by two approaches. The first, a phenomenological model based on physically realizable rational functions, enables the extraction of multiple resonance frequencies and widths for coupled resonators without explicit specification of the circuit network. In the second, an ABCD-matrix representation for the distributed transmission line circuit is used to model the observed response from the characteristic impedance and propagation constant. When used in conjunction with electromagnetic simulations, the kinetic inductance fraction can be determined with this method with an accuracy of 2\%. Datasets for superconducting microstrip and coplanar-waveguide resonator devices were investigated and a recovery within 1\% of the observed complex transmission amplitude was achieved with both analysis approaches. The experimental configuration used in microwave characterization of the devices and self-consistent constraints for the electromagnetic constitutive relations for parameter extraction are also presented.
\end{abstract}


\maketitle

\section{Introduction}
\label{sec:introduction}

Recent advances in astrophysical instrumentation have introduced superconducting microwave resonators for readout and the detection of infrared light.~\cite{NIKA, MUSIC, MAKO, Mazin} An essential element of the successful development of cameras employing these resonant structures is to accurately and efficiently characterize their properties. More broadly, precision measurements of quality factors $Q$, of superconducting resonators are desired in the context of microwave kinetic inductance detectors (MKIDs) device physics,~\cite{Zmuidzinas} quantum computing~\cite{Vijay} and the determination of low-temperature material properties.~\cite{Krupka,Oates} At sub-kelvin temperatures the quality factors are generally determined by measuring and fitting the transmission data over the relatively narrow bandwidth defined by the response of a single resonator.~\cite{Petersan} In this setting the fidelity of the observed response can be limited by the intervening instrumentation between the device under test and the measurement reference plane.

In physically modeling the electrical response of these circuits, a lumped-element approximation can be applied when the electrical size of the elements and their interconnections is small compared to a wavelength. This consideration ensures that the structures are not self-resonant. More generally, a transmission line approximation enables one to treat distributed elements with one dimension greater than a fraction of a wavelength in scale. However, when the electromagnetic fields and phases within the circuit elements are uniform and uncoupled, the lumped and distributed approaches are identical.~\cite{Bahl} It is not uncommon for ``lumped elements'' to have a significant internal phase delay in practical microwave devices. For ``capacitive'' elements synthesized from electrically small transmission line lengths this effect can be small, but ``inductive'' elements inherently possess a finite internal phase delay by design. The utility of this iconic lumped picture primarily resides in its conceptually simplicity as opposed to the end accuracy and fidelity of the physical representation.

This being said, analytical methods based on lumped-element approximations are commonly used to analyze scattering parameters of resonators and compute their $Q$ factors.~\cite{Gao,Khalil,Geerlings,Megrant,Deng,Swenson}
These expressions are derived from $LCR$ circuit representations and are based on the assertion of weak coupling between resonators to simplify the analysis. These formulations employ simplified functional forms for the transmission, $S_{21}$, and are suited for approximating the response near resonance for high-quality-factor lumped circuits.

A detailed review of fitting methods is offered in Ref.~\onlinecite{Petersan}, where several methods are presented for fitting the $S_{21}$ magnitude to different models, neglecting the measured phase. These methods do not deal with the alterations that the data incur in the real measurement scenario, such as crosstalk between cables and/or coupling structures, the resonator coupling ports not being coincident with the reference plane of the measurements, and the presence of and coupling to nearby resonators.
Many of the methods described that attempt to correct for these effects on $S_{21}$ remain sensitive to the details of fitting in the complex plane.~\cite{Petersan,Ma,Snortland}
In this work we strive to address these issues by using a transmission line representation to simultaneously analyze multiple coupled resonators and present a calibration methodology to remove non-ideal instrument artifacts.

The resonator's transmission amplitude and phase can be characterized with a vector network analyzer (VNA). Ideally, the VNA calibration reference plane or ``test port'' would be directly connected to the device under test (DUT); however, a cryogenic setting typically necessitates the use of additional transmission line structures to realize a thermal break between 300~K and the cold stage (e.g., 0.3~K in this work, see Fig.~\ref{fig:VNA_Measurement_Block_Diagram}). Other ancillary microwave components -- directional couplers, circulators, isolators, amplifiers, attenuators, and associated interconnecting cables -- may be required to appropriately prepare, excite, read out, and bias the device of interest. The phase velocities and dimensions of such components change in cooling from room temperature and therefore affect the observed response at the device calibration plane. These intervening components between the instrument and the device calibration plane influence the complex gain and need to be appropriately accounted for in the interpretation of the measured response.

Thermal and practical hardware constraints can pose significant but achievable calibration challenges at cryogenic temperatures. This can impact the data reduction approach and examples in the literature exist where complex measurements have been reduced to scalar quantities in the extraction of cryogenic resonator parameters.~\cite{Petersan} This is neither necessary nor desirable given that the scattering parameter phase contains important information which can be used to constrain the physical model of the system.

Ideally, a scheme is desired that places the reference plane of the measurement at the DUT which is identical to that used during calibration. To approximate this goal, measurements of the calibration standards on multiple cool-downs could be used; however, this places constraints on the VNA stability and cryogenic instrument repeatability, which at best are challenging, if not impractical, to achieve. To alleviate these concerns one can place the necessary standards in the cooled test environment and switch between them. Variations on this theme can be found in microwave probe stations, where the sample and calibration standards are cooled to 4~K by a continuous liquid helium flow or a closed-cycle refrigeration system.~\cite{Russell} This approach has been demonstrated at millikelvin temperatures but necessitates the availability of a low-power, high-performance microwave switch in the desired test band.~\cite{Ranzani} This three-standard Thru-Reflect-Line calibration approach was expanded to include error correction for \textit{in-situ} superconducting microwave resonator characterization.~\cite{Yeh}

For our metrology application, the measurement of the circuit's complex transmission, $S_{21}$, is needed. A flexible and \textit{in-situ} calibration procedure is desired for multiple resonators over a frequency range wide compared to the resonator line width and inter-resonator spacing. To explain the approach presented here, the reader is guided through a detailed description of the experimental set-up (Section~\ref{sec:experiment}), the motivation behind the transmission measurement metrology (Section~\ref{sec:metrology}), the calibration process (Section~\ref{sec:modeling}), and the derivation of the resonator parameters for two differing device datasets (Section~\ref{sec:ResonatorModel}). Finally, an ABCD-matrix fitting approach is used in extracting transmission line materials properties and is described in Section~\ref{sec:validation}.

\section{Experimental set-up}
\label{sec:experiment}
\begin{figure}[htbp]
	\centering
		\includegraphics[width=0.50\textwidth]{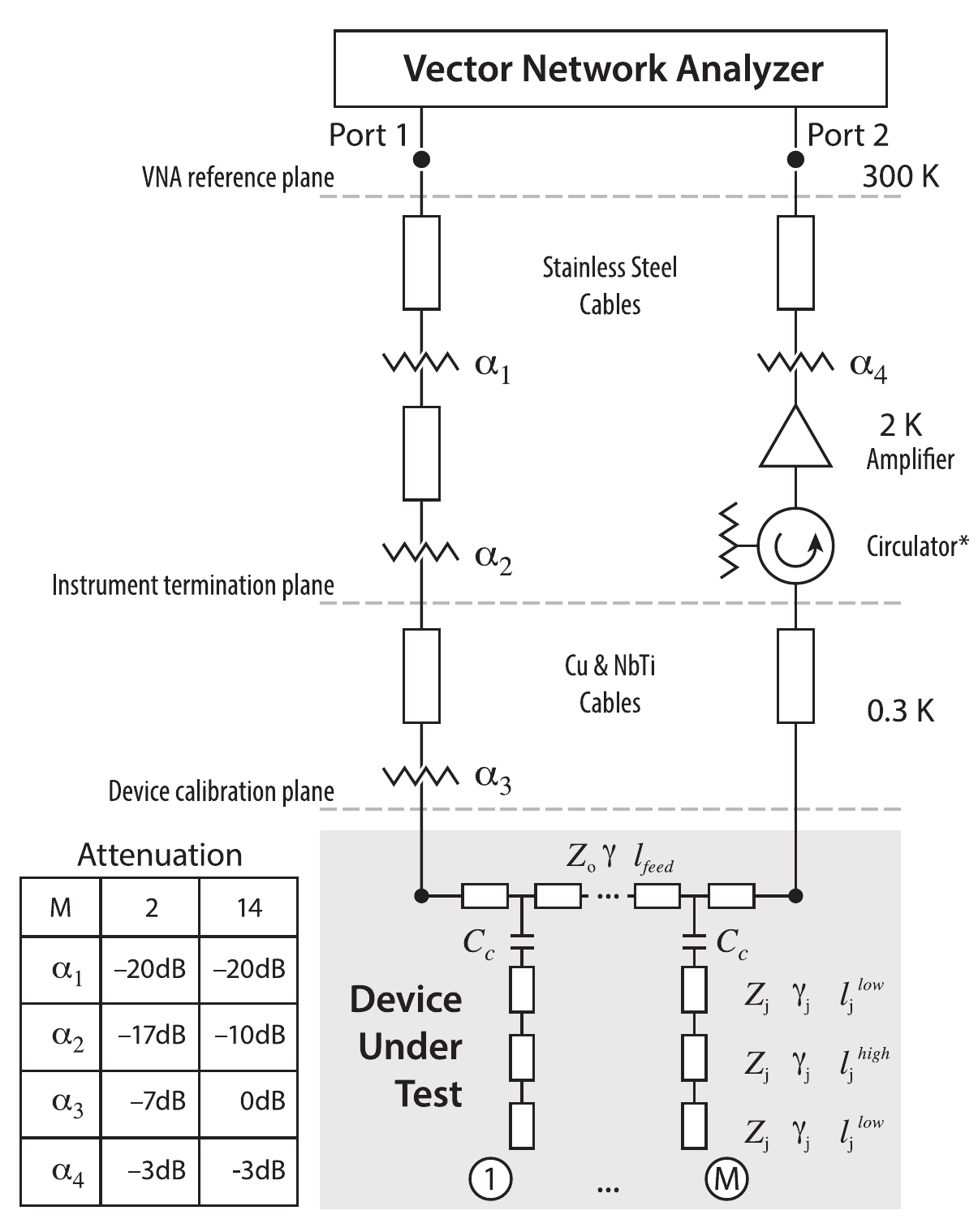}
	\caption{Schematic of the experimental apparatus and devices under test. The resonators are coupled to a CPW feedline through coupling capacitors and their response is observed through a VNA connected to the DUT. Summary of the attenuators employed in each measurement can be found in the table. The circulator indicated by a * was only present for the 2-resonator test configuration.}
	\label{fig:VNA_Measurement_Block_Diagram}
\end{figure}

We present datasets for the detailed characterization of two devices, the first with 2 resonators made on microstrip transmission lines and the second with 14 resonators in a coplanar waveguide (CPW) configuration. The experimental apparatus used for the measurements is depicted in Fig.~\ref{fig:VNA_Measurement_Block_Diagram}. We describe in detail the first device to demonstrate the technique and then report on the second device which contains multiple coupled resonators of varying quality factors as a demonstration of the algorithm's inherent robustness. 

\begin{figure*}[htbp]
	\centering
	\begin{subfigure}[b]{.30\textwidth}
		\includegraphics[width=1.00\textwidth]{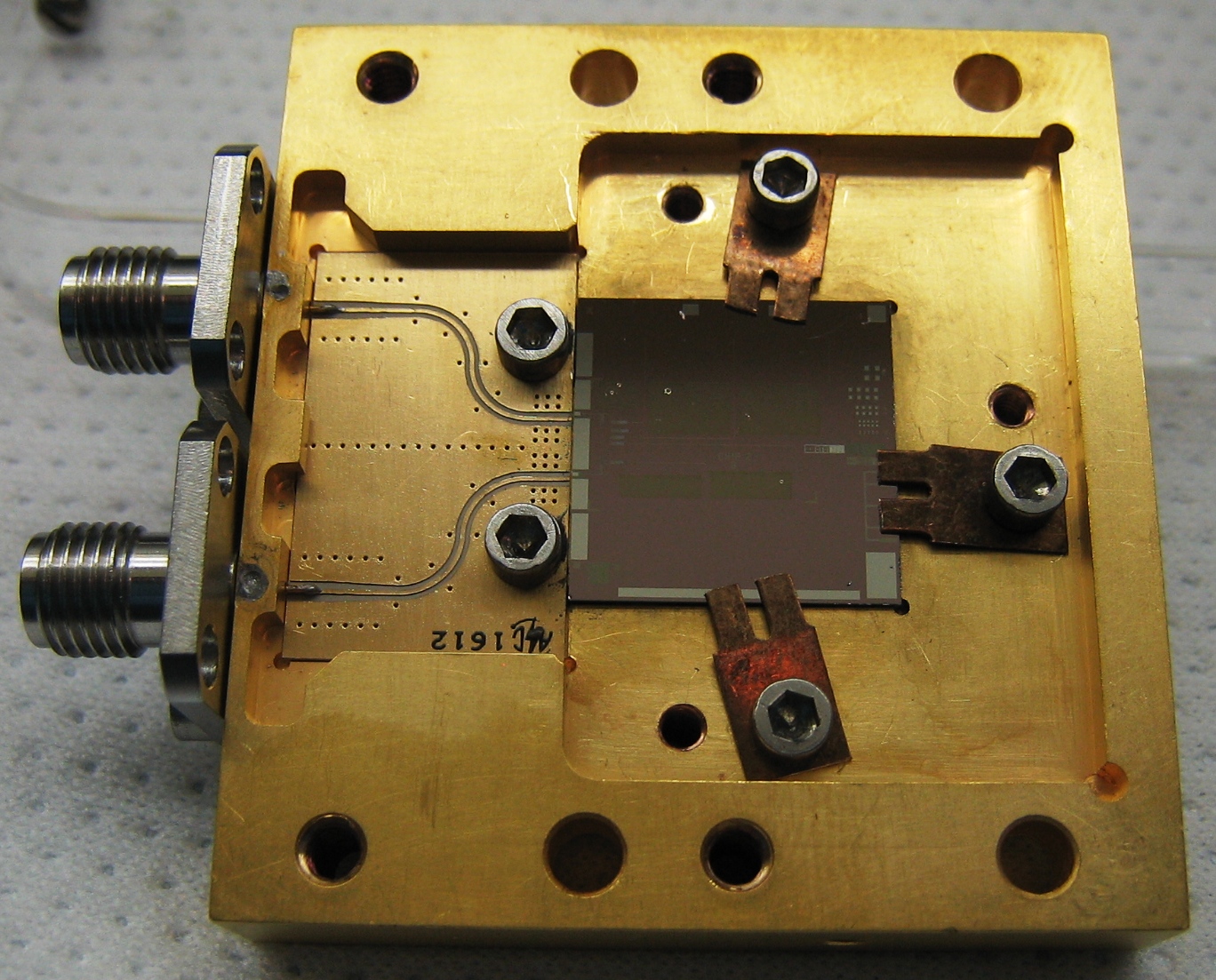}
		\vspace{10pt}
		\caption{Resonator chip, printed circuit CPW transition board, and SMA connector interface mounted in the test package. The package cover has been removed for clarity.}
		\label{fig:Chip on board}
	\end{subfigure}
	\qquad
	\begin{subfigure}[b]{.65\textwidth}
		\includegraphics[width=1.00\textwidth]{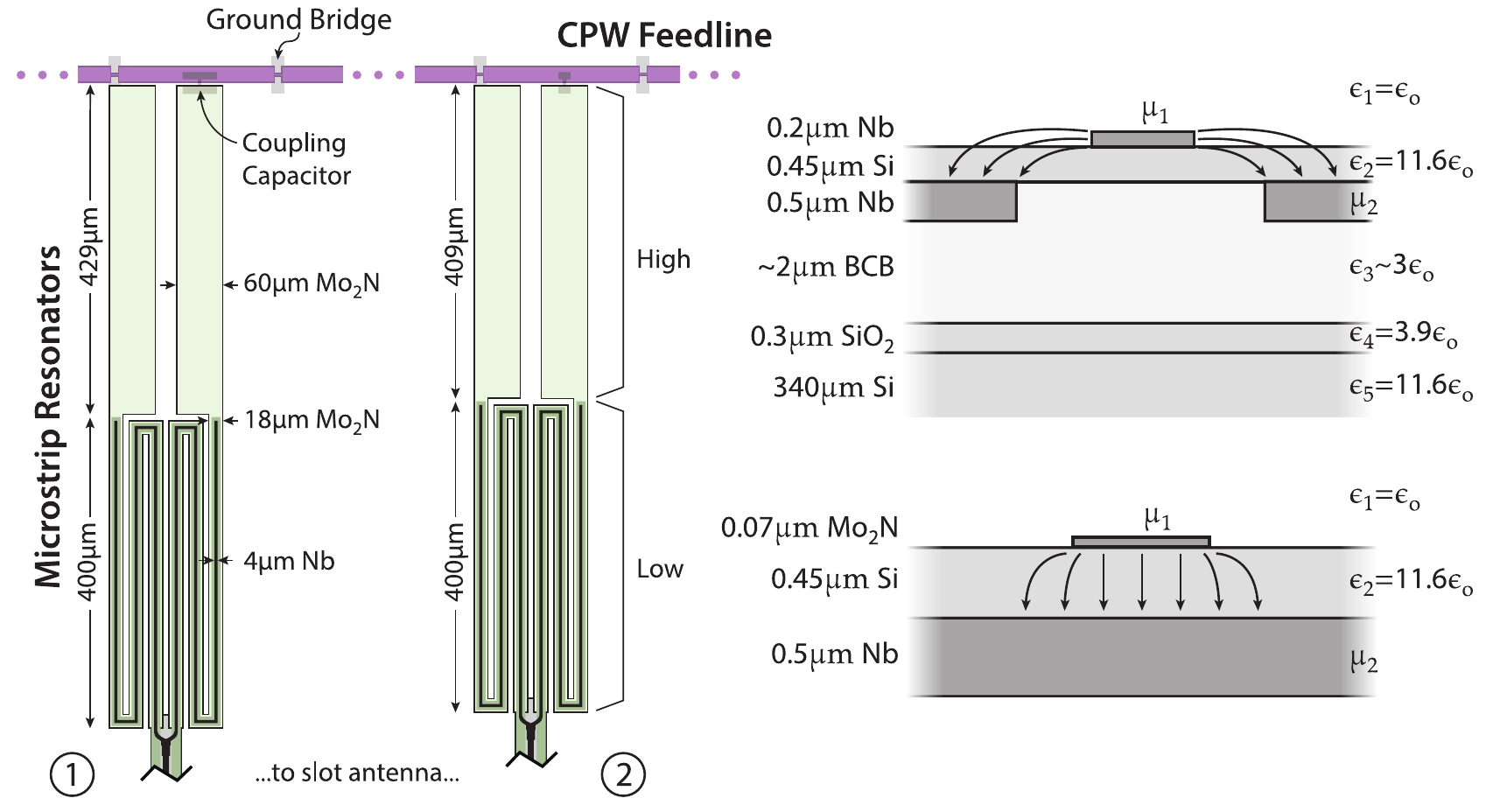}
		\caption{Left: Layout for the 2-resonator chip. The stepped-impedance resonators are coupled to a CPW feedline and are realized from low- and high-impedance microstrip transmission lines. The ``ground bridge'' prevents excitation of the asymmetric slotline mode on the CPW feedline structure. The 4-$\mu$m Nb microstrip line uses SiO$_{\text{2}}$ as a dielectric insulator (0.1-$\mu$m thick) from the Mo$_2$N layer. This transmission line structure is not explicitly used at microwave frequencies but enables millimeter wave coupling to a dual-slot antenna.~\cite{Patel} Right: A simplified cross-sectional view of the coplanar waveguide (top) and microstrip transmission line (bottom) geometries used in modeling the electromagnetic response. The permittivity for the dielectric layers and the permeability for the lines are indicated. The arrows indicate the electric field and its dominant modal symmetry. The microwave dielectric substrate (0.450$\pm$0.025-$\mu$m thick, 100-mm diameter) is monocrystalline silicon.~\cite{Soitec}}
		\label{fig:Resonators}
	\end{subfigure}
	\caption{Details of the 2-resonator device and packaging}
\end{figure*}

Figure~\ref{fig:Chip on board} shows the packaged 2-resonator test device. Here, a superconducting niobium (Nb) CPW feedline is coupled to two microstrip stepped-impedance superconducting molybdenum-nitride (Mo$_2$N) resonators.~\cite{Patel} This device was mounted inside a gold-plated copper (Cu) package. Aluminum wire bonds were used to connect the superconducting Nb CPW feedline to a Cu CPW line on a printed circuit transition board. These CPW lines were connected to Sub-Miniature version A (SMA) coaxial connectors whose center-pins were soldered to the printed circuit board. The packaged device was attached to the cryostat cold stage and cooled to 330~mK in a Helium-3 cryostat.

A series of coaxial cables and components routed the device input and output lines from 330~mK to an Agilent N5242A PNA-X network analyzer at room temperature. A Weinschel Model-4M SMA attenuator ($\alpha_3\simeq7$\,dB) with a measured return loss $\geq$ 20~dB from 2 to 4~GHz at 2~K was placed directly at the input of the device package to provide a matched termination for the test package input. Improved signal-to-noise can be achieved through the use of impedance-matched thermal blocking filters~\cite{Wollack} for realization of the terminations for the DUT. This configuration would also be desirable or necessary for the characterization of devices with low saturation power. The output signal was routed via copper and superconducting niobium-titanium (NbTi) coaxial cables through a CTD1229K PamTech circulator, which was mounted on the 2-K stage of the cryostat. This circulator operates with a return loss $\geq$~20~dB between 2.5-4~GHz. From the isolator the output signal was routed to an AF~S3-02000400-08-CR-4 MITEQ amplifier mounted on the 2-K stage.

\section{Measurement Metrology Motivation}
\label{sec:metrology}
We define the response at the ``VNA reference plane'' as observed through a microwave path to the ``device calibration plane'' as the instrument baseline (Fig.~\ref{fig:VNA_Measurement_Block_Diagram}). In a perfectly impedance-matched and lossless instrument, the magnitude of the complex transmission amplitude, $S_{21}$, for a thru connection would be unity. In practice, the instrument baseline has frequency structure and amplitude set by the details and quality of the intervening transmission line components. These instrumental artifacts need to be addressed in calibration to provide an unbiased measurement of the device response.

A Short-Open-Line-Thru (SOLT) calibration was used to calibrate the PNA-X in 3.5-mm coax at the VNA reference plane defined in Fig.~\ref{fig:VNA_Measurement_Block_Diagram}. After recording the complex transmission measurements, an \textit{in-situ} calibration process was used in order to move the reference plane to the device calibration plane. In the calibration process adopted here we assume that the test device sees a matched termination. For the 2-resonator data this condition is realized by a 7-dB attenuator and isolator. For the baseline removal approach described here to be successful, the response of the attenuators, transmission line media and amplifiers used in the system should be smooth functions over the spectral band of interest relative to the test-device frequency response.

It is notable that in this implementation the feed transmission line structure in the device under test plays a dual role: it allows excitation and read-out of the sharply peaked circuit response of interest, and off resonance serves as a controlled impedance reference (thru line) standard to specify the system's gain as a function of frequency. The underlying approach explored in this work for baseline removal has its roots in a variety of analogous instruments such as frequency-swept reflectometry~\cite{Somlo} and spectral baseline correction in radio astronomy.~\cite{Allen,Goldsmith2} In particular, the differential reflectometer configuration used for cavity characterization~\cite{RollWilkinson} provides insight in this context. In this method, the baseline is explicitly removed on resonance by comparing the response of a waveguide cavity with that of a short-circuit via the difference port of a magic-T power divider. Adjustment of the short position in this bridge reflectometer circuit reveals the interaction between the resonator and the baseline shape. More importantly, this points to the role of the resonator phase and baseline shape on the observed scattering parameters at resonance.

\section{VNA Transmission Data Calibration}
\label{sec:modeling}
For a single resonator or multiple well-separated resonators, the response can be well reproduced by a Lorentzian line shape. In the presence of multiple resonances, their mutual interaction as well as the interaction with the continuum can result in a Fano spectral response.~\cite{Fano1,Fano2,Marquezini,Singh,Giannini} This effect can also be experimentally observed as an interaction between the resonators with the relatively broad Fabry-Perot resonances resulting from standing waves in the system. Such reflections produce the dominant spectral variation in the observed instrument baseline and can be mitigated by minimizing the transmission line lengths and suitably terminating the reflections with matched attenuators and circulators at the instrument termination plane. In calibrating the spectra, an unbiased removal of these artifacts is desired.

\begin{figure*}[htbp]
	\includegraphics[width=0.90\textwidth]{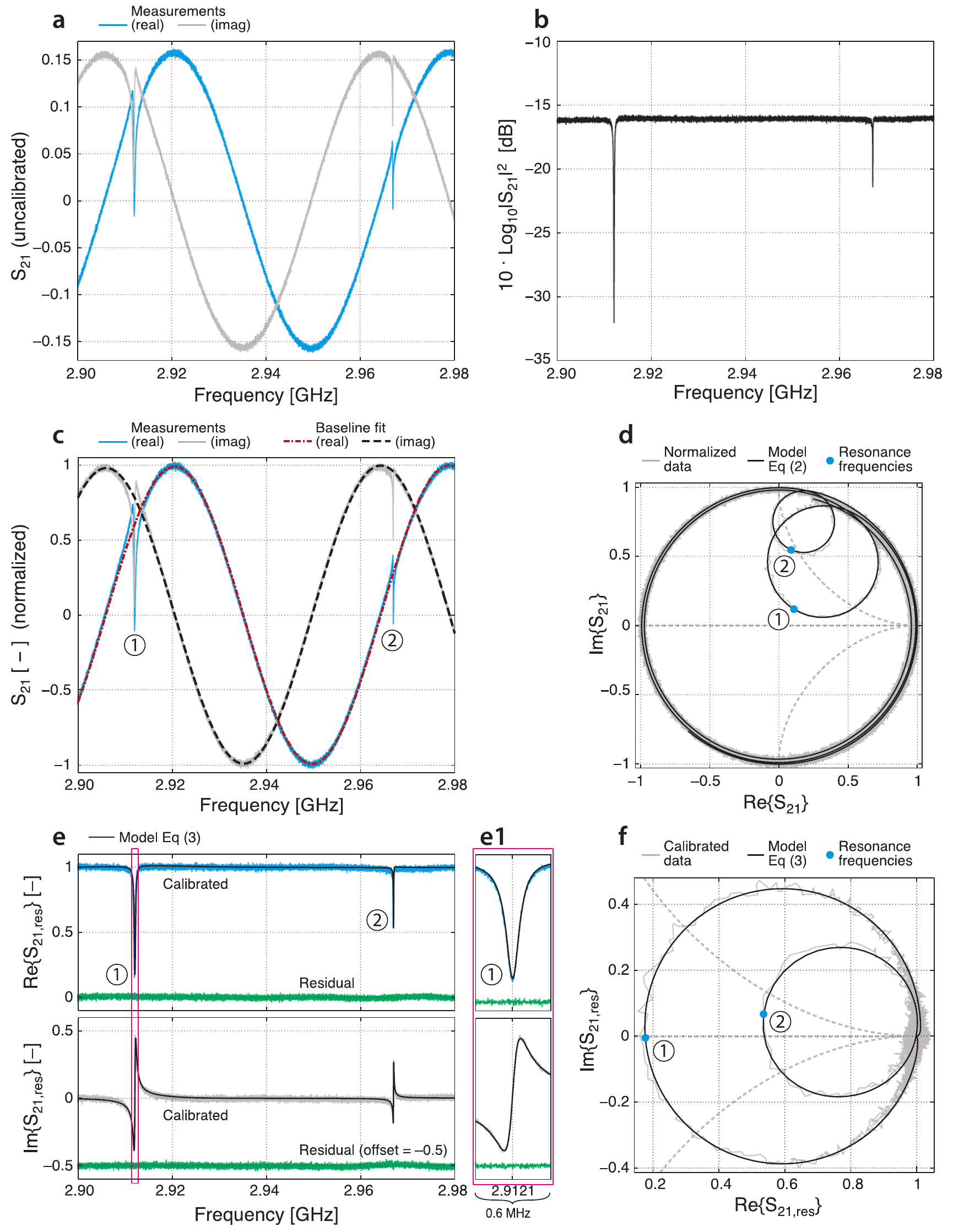}
	\caption{Measurement calibration overview. a)~Complex $S_{21}$ data for the DUT as observed at the VNA reference plane. b)~From the perspective of the DUT, the magnitude of the response is uncalibrated due to the amplifiers, attenuators, and other microwave elements in the system. c)~The transmission spectrum is normalized by forcing its amplitude to be equal to unity far from the resonator responses. d)~The normalized transmission data are shown on a Smith chart. The baseline variation with frequency can be seen as a deviation in the $S_{21}$ from unity far from the resonator response. e)~After removing the complex baseline and specifying the device-calibration-plane phase, the calibrated data are used to fit an analytical model, and the residuals are shown. f)~The Smith chart shows the two resonance loops going through (1,0) after calibration.}
	\label{fig:S21-Chart_2-resonator}
\end{figure*}

As an example of the calibration process, we use the data from the CPW feedline coupled to the two molybdenum-nitride (Mo$_2$N) resonators. Figure~\ref{fig:S21-Chart_2-resonator}a shows the real and imaginary components of the transmission, $S_{21}$, at the VNA reference plane as a function of frequency. It can be seen that the characteristic scale of the baseline variations is much larger than the resonator response in frequency. To calibrate the VNA data \textit{in-situ}, the following steps are performed:
\begin{enumerate}
	\item fit of the complex baseline (Fig.~\ref{fig:S21-Chart_2-resonator}c);
	\item	normalization of the transmission's real and imaginary parts (Fig.~\ref{fig:S21-Chart_2-resonator}c-d);
	\item correction for variations in gain and relocation of the reference plane at the DUT by ratioing out the complex baseline fit (Fig.~\ref{fig:S21-Chart_2-resonator}e-f).
\end{enumerate}

In general, the baseline can have an unpredictable shape as determined by the details of the reflections occurring throughout the instrument and its components.
The complex baseline, $S_{21,bas}$, was modeled analytically and a Fourier series was found to be a convenient and physically motivated representation of its response:
\begin{equation}
	S_{21,bas}=\sum_{j=1}^n A_j \cdot \exp{\left(i d_j \omega \right)}.
	\label{eq:baseline}
\end{equation}
In Eq.~\eqref{eq:baseline} $A_j$ is a complex coefficient, $d_j$ a time delay and $\omega$ the angular frequency. In the case presented, $n=4$ terms were found to be sufficient to adequately sample the baseline and appropriately constrain its properties. A fit to the 30,000 measured data points resulted in a reduced $\chi^2=0.9973$. Differences in experimental configuration or changes in the desired calibration spectral range may lead to alternative forms for Eq.~\eqref{eq:baseline}.

The second step in data calibration consists of uniquely specifying the complex gain amplitude between the VNA reference plane and the device calibration plane so that the $S_{21}$ real and imaginary components in Fig.~\ref{fig:S21-Chart_2-resonator}a lie between $\pm1$. This normalization factor was found to be approximately 6.32 by forcing the transmission amplitude to be equal to unity far from the resonator response. The normalized measurements and modeled data are presented in the Smith chart in Fig.~\ref{fig:S21-Chart_2-resonator}d.

Finally, the complex baseline, $S_{21,bas}$, was removed from the data through Eq.~\eqref{eq:baseline} to eliminate the influence of reflections and move the reference plane to the DUT. By looking at the topology of the instrument, the error matrices introduced by the connections between the reference plane and the device under test in the cryostat are in series. From the properties of signal flow graphs,~\cite{Somlo,Pozar} the uncalibrated transmission coefficient, $S_{21}$, at the VNA reference plane is therefore the product of the following scattering matrix elements:
\begin{equation}
	S_{21} = S_{21,bas} \cdot S_{21,res},
	\label{eq:model}
\end{equation}
where $S_{21,res}$ refers to the desired calibrated response of the resonator device under test.

\section{Phenomenological resonator model}
\label{sec:ResonatorModel}
With the baseline correction applied, the calibrated data (Figure~\ref{fig:S21-Chart_2-resonator}e) were modeled in order to extract the characteristic resonant frequencies and line widths. The feedline and resonator model for the packaged device can be represented as a realizable causal filter as follows:
\begin{equation}
	S_{21,res} = 1 + \sum_{j=1}^M \frac{a_{0,j} + a_{1,j}x_j + a_{2,j}x_j^2 + \ldots}{1 + i x_j + b_{2,j}x_j^2 + \ldots},
	\label{eq:resonators}
\end{equation}
where, for each $j=1,...,M$ ($M$ being the total number of resonators), $x_j=Q_{tot,j} \cdot \left(\omega/\omega_{o,j}-\omega_{o,j}/\omega\right)$ and $Q_{tot,j}$ is the total loaded quality factor defined as $Q_{tot,j}=\omega_{o,j}/\Gamma_j$.
For each resonator, therefore, the fitting parameters are: the $a_{k,j}$ and $b_{k,j}$ complex coefficients ($k=0,1,2,...$, $j=1,...,M$), the resonance frequency $\omega_{o,j}$, and the full width at half power $\Gamma_j$. Here we use $k=2$ in specifying the order of the polynomial.

The functional form of Eq.~\eqref{eq:resonators} exhibits several features. First, its second-order terms in both the numerators and denominators allow for reproduction of the resonator responses, thereby enabling the representation of a physically realizable, distributed circuit network.~\cite{Zemanian} Second, its causality is assured by the degree of the numerator being not greater than that of the denominator, which allows the functional form to satisfy the Kramers-Kronig relations.~\cite{Toll} Far removed from the resonators, $S_{21,res}\rightarrow 1+i\,0$ and the transmission represents an ideal thru line. It was verified that our second-order model approached this limit without additional terms to avoid increasing the number of parameters and the computational effort. Finally, this functional form enables one to simultaneously fit any number of resonators while formally taking into account their physical interactions. 

Rewriting $x_j$ in the equivalent form, $x_j = Q_{tot} \cdot (\omega_o/\omega_j)( \omega_j/\omega_o + 1) \cdot (\omega_j/\omega_o - 1)$, reveals a commonly employed approximation,  $x_j \approx 2 Q_{tot} \cdot (\omega_j/\omega_o - 1)$, near resonance. This form can be viewed as an asymptotic expansion in powers of $1/Q_{tot}$ which modifies the position of the resonant frequency in the complex plane.~\cite{Jackson} When Eq.~\eqref{eq:resonators} is evaluated in this limit the expressions reduce to the first-order lumped-circuit treatments described in the literature for a single resonator in transmission.~\cite{Gao,Khalil,Probst} However, we find that to faithfully reproduce the interactions between resonators over a wide spectral bandwidth the full functional form presented here is required.

\subsection{Analysis of the 2-resonator dataset}
\label{sec:2res}
Equation~\eqref{eq:resonators} was used to fit the calibrated data by means of a least-squares curve fitting routine based on a trust-region reflective Newton method.~\cite{Coleman1, Coleman2} Constraints were given in the form of global lower and upper bounds for each $\Gamma>0$ and $\omega_o$. The starting guess for each $\Gamma$ was chosen close to the actual value, which can be readily determined by estimating each resonator's spectral width. The initial guess for the resonance frequency, on the other hand, corresponds to the frequency values where the observed $S_{21}$ has minima. The results are shown in Fig.~\ref{fig:S21-Chart_2-resonator}e (black line) for both components. The residuals on the real and imaginary parts are comparable to the normally-distributed noise level in the measured data with a reduced $\chi^2=0.9987$ and a standard deviation of $\sigma=0.0155$. A Smith chart is provided in Fig.~\ref{fig:S21-Chart_2-resonator}f, which shows that the two circles touch each other in $(1,0)$ with a relative rotation of about $7^\circ$ caused by the phase delay in the feedline length between the two resonators. Because of the noise in the measured data and the residual systematic errors in the baseline fit, the data around $(1,0)$ converge to this point within a radius of $0.04$. The estimated values for the fitting parameters of each resonator are summarized in Table~\ref{tab:fit}.

\begin{table*}[!t]
  \caption{Parameter summary for the 2-resonator analytical model\footnote{In the analysis double precision (16 digits) was used; however, for clarity here we only show the leading digits.}}
  \begin{ruledtabular}
	\begin{tabular}{ccccccc}
		$j$   & $a_{0,j}$ & $a_{1,j}$ & $a_{2,j}$ & $b_{2,j}$ & $\omega_{o,j}/(2\pi)$ & $\Gamma_j/(2\pi)$ \\
		$[-]$ & $[-]$     & $[-]$     & $[-]$     & $[-]$     & $[$GHz$]$             & $[$kHz$]$  \\
		\hline
		$1$  & $-0.7345+i\,0.1029$ & $-0.0440+i\,0.0879$ & $-0.0008-i\,0.0009$ & $-0.0120-i\,0.0048$  & $2.9121\ldots \pm  7\times10^{-9}$ & $139.02 \pm 0.01$ \\
		$2$  & $-0.5496-i\,0.0463$ & $-0.4737+i\,2.1868$ & $-2.1104-i\,2.4638$ & $30.0366-i\,13.2458$ & $2.9670\ldots \pm 60\times10^{-9}$ & $ 50 \pm 2$ \\
	\end{tabular}
	\end{ruledtabular}
	\label{tab:fit}
\end{table*}
The internal and coupling $Q$ factors were calculated using the following equations:~\cite{Kajfez,Canos}
\begin{eqnarray}
	Q_{i,j} = \frac{Q_{tot,j}}{1-D_j}, \\
	Q_{c,j} = \frac{Q_{tot,j}}{D_j},
	\label{eq:Qic}
\end{eqnarray}
where $D_j$ represents the diameter of the circle associated with each resonator (Fig.~\ref{fig:S21-Chart_2-resonator}f).
The $Q$-factor values are shown in Table~\ref{tab:Q}.
\begin{table}[htbp]
  \caption{$Q$ factors for the 2-resonator analytical model}
	\begin{ruledtabular}
	\begin{tabular}{cccc}
		 $j$  & $Q_{tot,j}$ & $Q_{i,j}$ & $Q_{c,j}$  \\
		$[-]$ & $[-]$       & $[-]$     & $[-]$      \\
		\hline
		$1$   & $20,948 \pm   1$ & $107,583 \pm   2$ & $ 26,013 \pm   1$ \\
		$2$   & $59,400 \pm 180$ & $109,500 \pm 330$ & $129,800 \pm 390$ \\
	\end{tabular}
	\end{ruledtabular}
	\label{tab:Q}
\end{table}

\subsection{Analysis of the 14-resonator dataset}
\label{sec:14res}
\begin{figure*}[htbp]
	\includegraphics[width=1.00\textwidth]{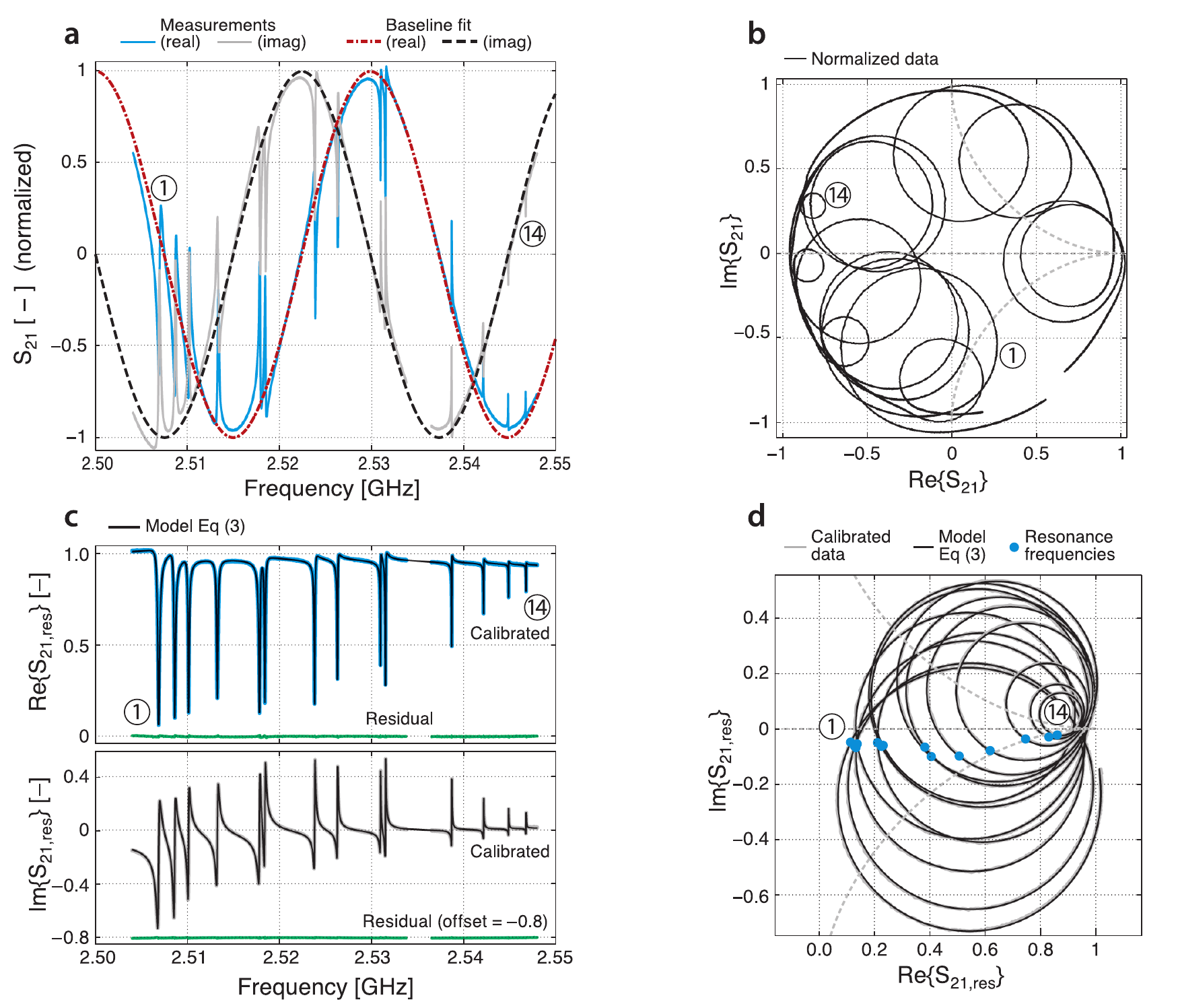}
	\caption{Results for the 14-resonator CPW dataset. a) Normalized transmission data. b) Smith chart representation of the normalized transmission data. c) Calibrated transmission data and analytical model. The achieved residuals are smaller than 1\%. d) Smith chart showing the calibrated 14-resonator data wrapping around (1,0). Note: The 14-resonator CPW dataset was acquired in two discrete sections and a small gap in the measured transmission spectra is present at $\approx2.535\,$GHz.}
	\label{fig:S21-Chart_14-resonator}
\end{figure*}

A 14-resonator CPW dataset was also studied (see Fig.~\ref{fig:S21-Chart_14-resonator}a-b). In this example, the resonators span a spectral range of about 45~MHz with two pairs of resonators strongly interacting with each other, namely resonators 5-6 and 9-10. The analysis of this dataset was challenging because, for high values of $Q_{tot}$, the denominator of Eq.~\eqref{eq:resonators} approaches an indeterminate form of the type $[0\cdot\infty]$, which can result in numerical instability. In addition, the parameters of interest span 6 orders of magnitude leading to an ill-conditioned Hessian matrix for the system. 

Equation~\eqref{eq:resonators} was used to fit the calibrated data by means of a least-squares curve fitting routine based on a trust-region reflective Newton method with a diagonal preconditioning of the conjugate gradient. To help the algorithm with convergence, all $\Gamma_j$ and $\omega_{o,j}\;(j=1,\ldots,14)$ were provided with lower and upper bounds. Namely, $0<\Gamma_j\leq\Gamma_{max}$ and $\omega_{min}\leq\omega_{o,j}\leq\omega_{o,max}$ for $j=1,\ldots,14$.
Here, $\Gamma_{max}=10$ MHz, well above the largest $\Gamma_1$ ($\approx1.6$ MHz). The minimum angular frequency is $\omega_{min}=2.5040$~GHz and the maximum resonance frequency, $\omega_{o,max}$, was set not to exceed a distance of $10\times\Gamma_{14}$ from the last resonance frequency. For the highest-$Q$ resonators (i.e., the highest frequencies in the recorded transmission spectra) the model is more sensitive to changes in $\omega_{o,j}$. This choice for the bounds helped the algorithm in identifying solutions within the desired range of parameters and converged after $<300$ iterations.

The model represented by Eq.~\eqref{eq:resonators} recovers the calibrated data within an accuracy of $<1\%$ (Fig.~\ref{fig:S21-Chart_14-resonator}c-d) with a reduced $\chi^2=0.9985$ and a standard deviation $\sigma=0.0027$, thus proving to be a robust method to analyze numerous, strongly coupled resonators. In this dataset, the range of quality factors spans more than three orders of magnitude, e.g., $2,300 < Q_{c} < 630,000$. A particular challenge noted in this existing dataset was that the observed spectral range was suboptimal. From an experimental design perspective, the phase variation of the DUT and the baseline need to be appropriately sampled and to possess differing spectral signatures to enable independent reconstruction from the measured dataset. By using the modeled response as noiseless input to the fitting algorithm, it was found that the spectral range of this dataset should have been extended by $30\%$ to minimize the calibration error with this algorithm. This increase in spectral range would enable the responses of the resonator wings and the baseline to be sufficiently decoupled to achieve an unbiased amplitude calibration.

\section{ABCD-matrix Model}
\label{sec:validation}
The phenomenological resonator model based on Eq.~\eqref{eq:resonators} and discussed in the previous section is useful and sufficient if one needs to know the center frequencies and the resonators' $Q$ factors. Clearly, there are many quasi-TEM structures such as microstrip, CPW, or other transmission line types which can lead to the observed response. Since the phenomenological resonator model is blind to the details of the realization, to probe the internal structure of the circuit a distributed transmission line representation needs to be employed. Here, the impedance, propagation constant, and lengths are used to characterize the circuit response in an ABCD or chain-matrix formulation. In this process, specifically, the impedance and propagation constant are used as parameters to fit the transmission data.

Although coupling to higher-order modes (e.g., losses to radiation, surface waves, and parasitic coupling between elements) can influence the underlying network topology, in the formulation used here we assume single-mode propagation and rely upon the specification of the circuit network.
This process is related to ``de-embedding'' or the process of inferring the response of a device under test when electrical properties of the intervening structure (in this case, the feedline coupling structure) are known.~\cite{Bauer} Elegant implementations of such concepts in the context of scalar millimeter-wave circuit metrology are described in Refs.~\onlinecite{Vayonakis,Rebeiz}.

To model the 2-resonator device response below the instrument termination plane (Fig.~\ref{fig:VNA_Measurement_Block_Diagram}), 29 transmission line elements were used. For each line section the physical lengths, $l$, were known to a micron accuracy from the photolithographic mask, while the characteristic impedances, $Z_o$, and the propagation constant, $\gamma$, were simulated with HFSS (High Frequency Structure Simulator). For simplicity, the two coupling capacitors were modeled as lumped elements and their capacitances also simulated with HFSS ($C_{c_1} \approx 53\,$fF and $C_{c_2} \approx 23\,$fF). The detailed description of the ABCD-matrix model can be found in Appendix~\ref{sec:appendix}.

Substituting the numerical values of each transmission line element into Eq.~\eqref{eq:ABCD_TL} and cascading through Eq.~\eqref{eq:cascade} yields the ABCD parameters of the entire transmission line circuit. These parameters were used in Eq.~\eqref{eq:S21} to evaluate the modeled transmission response, $S_{21}$. A challenge of this method lies in the difficulty of fully specifying the elements of the system at cryogenic temperatures. To simplify the extent of the system to be characterized, the instrument calibration plane was chosen to approximate a matched termination as close as feasible to the packaged device under test.

In addition, as a cross-check of the baseline correction obtained by fitting, the elements leading to this instrumental response were computed with the ABCD-matrix model and agreed to within a $5\%$ accuracy with the measured transmission line coaxial-cable lengths given nominal literature values for the teflon dielectric. Experimental investigation revealed that the superconducting NbTi thermal break cable was the limiting element in the uncorrected baseline response. In fitting, it was found that the characteristic impedance of this line was $\approx60\,\Omega$. This section of cable was subsequently inspected and a gap in the teflon dielectric was noted to be consistent with these observations.

\subsection{Transmission line parameter extraction}
\label{sec:TL}
For the transmission line structures the characteristic impedance, $Z_o$, and propagation constant, $\gamma$, are functions of the effective permittivity, $\varepsilon_{r,\mbox{\scriptsize{eff}}}$, and permeability, $\mu_{r,\mbox{\scriptsize{eff}}}$, in the medium. In particular, $Z_o$ is a function of the transmission line geometry proportional to the relative wave impedance in the medium, $Z_n \equiv (\mu_{r,\text{eff}}/\varepsilon_{r,\text{eff}})^{1/2}$, and $\gamma$ is defined such that:
\begin{equation}
	\gamma^2 \equiv - \left(\frac{\omega}{c}\right)^2 \cdot (\varepsilon_{r,\text{eff}} \cdot \mu_{r,\text{eff}})
	= - \left(\frac{\omega}{c}\right)^2 \cdot n_{\text{eff}}^2,
	\label{eq:betal}
\end{equation}
where $\omega$ is the angular frequency, $c$ the speed of light in vacuum, and $n_{\text{eff}}$ is the effective index in the medium. For microstrip, the effective permittivity and permeability are explicitly defined as:
\begin{equation}
	\varepsilon_{r,\text{eff}} \equiv \frac{C(\varepsilon_1,\varepsilon_2)}{C(\varepsilon_o,\varepsilon_o)},
	\quad \mu_{r,\text{eff}} \equiv \frac{L(\mu_1,\mu_2)}{L(\mu_o,\mu_o)},
	\label{eq:consec_relation}
\end{equation}
where the capacitance, $C$, and the inductance, $L$, per unit length are measures of the electromagnetic energy stored in the transmission line.~\cite{Harrington,Hoffmann} To find the effective constitutive relations, these functions are evaluated in the presence and absence of the dielectric and superconducting media, respectively. Here, $\varepsilon_2$ refers to the monocrystalline silicon substrate, $\varepsilon_1$ to the superstrate dielectric, $\mu_1$ to the centerline, $\mu_2$ to the ground plane metallization, and $\varepsilon_o$ and $\mu_o$ are the permittivity and permeability of free space (see Fig.~\ref{fig:Resonators}). For other transmission line structures, analogous formulations of Eq.~\eqref{eq:consec_relation} can be defined for the constitutive parameters.

This parameterization results in $\varepsilon_{r,\text{eff}}$ and $\mu_{r,\text{eff}}$ becoming implicit functions of the line's materials properties and cross-sectional geometry. The imaginary component of $\mu_{r,\text{eff}}$ was taken as zero. An effective dielectric loss tangent $\approx1.8\times10^{-5}$ was observed in fitting the observed microstrip resonator spectra. This is consistent with the bound on the dielectric loss tangent, $<5\times10^{-5}$, at 2.2~K for a similar (1.45-$\mu$m-thick) silicon sample used as a Nb microstrip ring resonator at 4.7~GHz. As discussed in Ref.~\onlinecite{Datta}, upon cooling to cryogenic temperatures the bulk resistivity silicon ($\rho \ge 1\,\text{k}\Omega-\text{cm}$ for these samples) freezes out and the intrinsic loss mechanisms in silicon dominate at microwave frequencies. The achievable thicknesses and high uniformity ($\pm 0.013 \,\mu{\rm m}$) of the monocrystalline silicon wafers~\cite{Soitec} employed in this investigation enable its use as an ultra-low-loss controlled-impedance microwave substrate. The observed loss in the circuit can have contributions from the bulk properties of the silicon layers, the bisbenzocyclobutene (BCB) wafer bonding agent, two-level systems in the silicon oxide layer, and conversion to surface and radiation modes. In addition, residual coupling to the electromagnetic environment could contribute to the observed loss, in particular for higher $Q$ factors. Although a detailed breakdown of the contributions to the observed loss is possible by investigating the performance as a function of geometry and temperature, this was not attempted here. Thus, in ascribing the entire observed loss to the dielectric, the above-mentioned loss tangent represents a conservative upper bound to that of monocrystalline silicon.

In fitting, for each transmission line type (high or low) a global characteristic impedance and phase velocity were employed. The low-impedance lines have a width of $60\;\mu$m, whereas the high-impedance lines have a width of $18\;\mu$m. Electromagnetic simulations were performed with HFSS for the CPW and microstrip configurations shown in Fig.~\ref{fig:Resonators}. A study of the boundary conditions for the 18-$\mu$m line width is summarized in Table~\ref{tab:HFSS}, whose last column represents the configuration used in the device under test. To incorporate the effects of the kinetic inductance in the simulation, a surface inductance per square, $L_k^\square$, was introduced.~\cite{Kerr} Waveports matched to the transmission line impedance and an S-parameter amplitude convergence $\Delta S < 0.01$ were employed in the finite-element simulations. By comparing to analytical expressions for microstrip geometries, a conservative systematic error of $< 2\%$ was estimated for the simulated parameters derived from HFSS. Similar estimates arise from consideration of the magnitude of the constitutive relations derived from the simulations in the limit where the dielectric materials are replaced by free space and metallization layers are perfect-electric field boundaries.
\begin{table*}[htbp]
  \centering
  \caption{HFSS simulations -- Superconducting microstrip transmission line (18-$\mu$m line width) \footnote{The upper portion of the table contains the simulation input parameters and the lower portion the computed and derived results.}}
  	\begin{ruledtabular}
    \begin{tabular}{rllllr}
    Transmission line conductor, model definition  					& Perfect-E & Perfect-E & Nb & Mo$_2$N &  \\
    Line surface inductance, $L_k^\square$(line)					  & --  	& -- 	 	& 0.12  & 4.3  & [pH/$\square$] \\
    Line metallization thickness 														& --  	& --  	& 0.25  & 0.07  & [$\mu$m] \\
    Ground plane conductor, model definition  							& Perfect-E & Perfect-E & Nb & Nb &  \\
    Ground plane surface inductance, $L_k^\square$(ground)	& -- 		& --	  & 0.12  & 0.13  & [pH/$\square$] \\
    Ground plane metallization thickness 										& --	  & --	  & 0.50  & 0.50  & [$\mu$m] \\
    Superstrate relative permittivity, $\varepsilon_{1}/\varepsilon_{o}$	& 1.000  & 1.000  & 1.000  & 1.000  & [--] \\
    Substrate relative permittivity\footnote{For the simulations presented, the input dielectric function for silicon was specified as lossless for simplicity.}, $\varepsilon_{2}/\varepsilon_{o}$		& 1.000  & 11.55 & 11.55 & 11.55 & [--] \\  \hline
    Transmission line impedance, $Z_{o,\text{HFSS}}$  	& 8.5(0)   & 2.6(1)   & 3.1(1)  & 7.8(4)  & [$\Omega$] \\
    Relative wave impedance, $Z_n = (\mu_{r,\text{eff}}/\varepsilon_{r,\text{eff}})^{1/2}$		& 1.00(0)\footnote{The last significant digit computed in HFSS simulations is indicated with parentheses for selected entries. Deviation from zero is suggestive of the magnitude of the systematic error encountered in simulation of the transmission line structures.} & 0.307 & 0.366 & 0.922 & [--] \\
    Effective index, $n_{\text{eff}} = (\varepsilon_{r,\text{eff}} \cdot \mu_{r,\text{eff}})^{1/2}$ & 1.00(2) & 3.293 & 3.999 & 10.67 & [--] \\
    Effective relative permittivity, $\varepsilon_{r,\text{eff}} = n_{\text{eff}}/Z_n$ & 1.00(2) & 10.73 	 & 10.93 	& 11.57 	& [--] \\
    Effective relative permeability, $\mu_{r,\text{eff}} = n_{\text{eff}} \cdot Z_n$   & 1.00(2) & 1.0(1) & 1.463 	& 9.844 	& [--] \\
    Kinetic inductance fraction, $\alpha$ = 1-1/$\mu_{r,\text{eff}}$ 									 & 0.00(2) & 0.0(1) & 0.317 	& 0.898 	& [--] \\
    \end{tabular}
 	\end{ruledtabular}
  \label{tab:HFSS}
\end{table*}

To study the resonator circuit's response in detail, a fit of the $S_{21,res}$ data depicted in Fig.~\ref{fig:S21-Chart_2-resonator}e to Eq.~\eqref{eq:S21} was performed. The fitting parameters were $Z_o$, $n_{\text{eff}}$, and $C_c$ for each of the two resonators' transmission line sections. From $Z_o$, the medium's relative wave impedance, $Z_n$, can be determined as follows:
\begin{equation}
	Z_n \equiv \left( \frac{\mu_{r,\mbox{\scriptsize{eff}}}} {\varepsilon_{r,\mbox{\scriptsize{eff}}}} \right)^{1/2}
	= \frac{Z_o}{Z_{o,\mbox{\scriptsize{HFSS}}}} 
	\label{eq:Zn}
\end{equation}
for each of the transmission lines used in the resonator realization. Here, $Z_{o,\text{HFSS}}$ represents the transmission line characteristic impedance simulated using HFSS with perfect E-field boundaries for conductors and the permittivity of free space for dielectrics.

From Eq.~\eqref{eq:betal} and Eq.~\eqref{eq:Zn} it follows that the effective permittivity and permeability are related by
\begin{equation}
	\varepsilon_{r,\text{eff}} = n_{\text{eff}} / Z_n,
	\quad \mu_{r,\text{eff}} =   n_{\text{eff}} \cdot Z_n.
\end{equation}
These scaling relationships between $\varepsilon_{r,\text{eff}}$ and $\mu_{r,\text{eff}}$ were enforced by the algorithm during fitting. The resulting transmission response computed with this methodology [see Eq.~\eqref{eq:S21}] agrees with the measured data within 1\%, similarly to the accuracy previously found through the evaluation of Eq.~\eqref{eq:resonators}. This analysis approach is analogous to that implemented in Ref.~\onlinecite{Chuss} to explicitly link the transmission line parameters which depend upon the detailed sample geometry to the interaction of the fields in a homogeneous bulk material sample in an electromagnetically self-consistent manner. For example, a non-physical dielectric function would be encountered if $\mu_{r,\text{eff}} \equiv 1$ was tacitly adopted for the superconducting material.~\cite{Mei-Liang,Shen} 

\begin{table}[htbp]
  \caption{Transmission line parameter extraction -- 2-resonator ABCD-matrix model}
	\begin{ruledtabular}
	\begin{tabular}{lcccc}
		 Line  	& $Z_{o,\text{HFSS}}$ & $Z_o$ 			& $n_{\text{eff}}$ & $\alpha$ \\
		 				& $[\Omega]$ 				 & $[\Omega]$ 	& $[-]$            & $[-]$ \\
		\hline
		1, low  & 2.7(0)\footnote{The last significant digit computed in HFSS simulations is indicated with parentheses.} &	$2.47\pm0.01$ & $ 10.697\pm0.571$ & $0.898\pm0.008$ \\
		1, high & 8.5(0) & $7.70\pm0.10$ & $11.143\pm0.336$ & $0.901\pm0.007$ \\
		2, low  & 2.7(0) & $2.47\pm0.01$ & $10.701\pm0.162$ & $0.898\pm0.004$ \\
		2, high & 8.5(0) & $7.70\pm0.03$ & $11.147\pm0.084$ & $0.901\pm0.003$ \\
	\end{tabular}
	\end{ruledtabular}
	\label{tab:alpha}
\end{table}

A parameter of interest for superconducting resonator-based detectors is the kinetic inductance fraction,~\cite{Porch} $\alpha$, which is defined as the ratio of the transmission line's kinetic inductance, $L(\mu_1,\mu_2)-L(\mu_o,\mu_o)$, relative to the total inductance, $L(\mu_1,\mu_2)$. With these definitions and through the use of Eq.~\eqref{eq:consec_relation}, one obtains:
\begin{eqnarray}
	\alpha &=& 1-\frac{1}{\mu_{r,\text{eff}}}.
	\label{eq:alpha}
\end{eqnarray}
The results found for $\alpha$ are in quantitative agreement with the values derived from the HFSS simulations (Table~\ref{tab:HFSS}) and can be found in Table~\ref{tab:alpha}.

\section{Conclusions}
\label{sec:conclusions}
This work presented a methodology to calibrate the transmission response of superconducting microwave resonators at cryogenic temperatures and described two methods to analyze the resulting dataset. The phenomenological and ABCD-matrix methods recovered the measured transmission data with the same 1\% level of accuracy, thus providing a numerical validation of the general approach. The derived $Q$ factors in either approach were statistically indistinguishable.

It is interesting to compare the number of parameters required by each method. The first analysis method employs a rational polynomial function of degree $k$ with $2k$ complex and 2 real parameters for each resonator. A second-order function with 6 parameters per resonator was shown to recover the measured data to within a 1\% accuracy. The second method, based upon an analysis of the circuit's distributed network, was implemented via cascaded ABCD matrices. Here, 2 complex and 3 real parameters were necessary to compute the ABCD matrices of each transmission line section used to represent the resonator structure and the feedline.
In addition, the line lengths were taken as known and the constraint between the wave impedance and the effective index was enforced in fitting. Broadly speaking the two approaches have comparable underlying complexity.

While the phenomenological model is useful in providing the values of the resonators' central frequencies and widths, the ABCD-matrix method provides insight into the circuit's internal structure when used in conjunction with electromagnetic simulation tools, to make an explicit linkage between the model and the underlying geometric details of the transmission media in use. This approach can be of particular value in extracting the line parameters required for precision circuit design. More importantly, the ABCD-matrix model naturally enables the treatment of distributed transmission line structures without the approximations typically employed in conventional lumped-circuit analysis. The accuracy of these analysis methods exceeds that of a simple lumped-circuit approximation over the frequency span and parameter range experimentally explored.

\begin{acknowledgments}
We acknowledge financial support from the NASA ROSES/APRA program and the Massachusetts Institute of Technology ``Arthur Gelb'' fellowship. We would like to thank Amil Patel for sample fabrication as well as David Chuss, Negar Ehsan, Omid Noroozian, Jack Sadleir, and Thomas Stevenson for helpful conversations and contributions to this work.
\end{acknowledgments}

\appendix
\section{}
\label{sec:appendix}
The ABCD matrix for a two-port network is defined in terms of the total voltages and currents as follows:~\cite{Goldsmith,Pozar}
\begin{equation}
	\left[ \begin{array}{c} V_{in} \\ I_{in} \end{array} \right] = \left[ \begin{array}{cc} A&B \\ C&D \end{array} \right] \left[ \begin{array}{c} V_{out} \\ I_{out} \end{array} \right].
	\label{eq:ABCD}
\end{equation}
The scattering parameter $S_{21}$, also known as the transmission amplitude, $t$, can be directly related to the ABCD parameters as:~\cite{Goldsmith}
\begin{equation}
	S_{21} = \frac{2Z_l}{A Z_l + B + C Z_l Z_s + D Z_s},
	\label{eq:S21}
\end{equation}
where $Z_l$ and $Z_s$ represent the load and source impedances, respectively.
The ABCD matrix of the cascade connection of multiple networks is equal to the product of the ABCD matrices representing the individual two-ports, that is
\begin{equation}
	\left[ \begin{array}{cc}A&B\\C&D\end{array} \right] = \prod_i \left[ \begin{array}{cc} A_i & B_i \\ C_i & D_i \end{array} \right]_i.
	\label{eq:cascade}
\end{equation}
Here, the subscript $i$ is used to indicate the different matrices as well as the ABCD parameters of each matrix.

In particular, the ABCD parameters of a two-port circuit represented by a transmission line are:~\cite{Goldsmith,Pozar}
\begin{equation}
	\left[ \begin{array}{cc}A&B\\C&D\end{array} \right]_{TL} = \left[ \begin{array}{cc}\cosh(\gamma l) & Z_o\cdot\sinh(\gamma l)\\ 1/Z_o\cdot\sinh(\gamma l) & \cosh(\gamma l) \end{array} \right].
	\label{eq:ABCD_TL}
\end{equation}
In Eq. \eqref{eq:ABCD_TL}, $Z_o$ is the characteristic impedance, $\gamma$ the propagation constant, and $l$ the line length. The determinant of this matrix is unity.

Each resonator is made of a low-, high-, and low-impedance transmission line connected to a coupling capacitor (Fig.~\ref{fig:Resonators}). The resonator's low-impedance transmission line section termination was modeled as an open-circuited line, i.e., $Z_{in,1} = Z_1 \coth{\gamma_1 l}_1$, while the high- and second low-impedance transmission lines were specified by~\cite{Pozar}
\begin{equation}
	Z_{in,j} = Z_{j}\frac{Z_{in,j-1}\cosh(\gamma_j l_j) + Z_j\sinh(\gamma_j l_j)}{Z_{in,j-1}\sinh(\gamma_j l_j) + Z_j\cosh(\gamma_j l_j)}, \quad j = 2,3.
\end{equation}
Finally, the total impedance as seen through the coupling capacitors is $Z_{tot} = 1/(i\omega C_c)+Z_{in,3}$. Here, the index $j$ specifies the internal lines in the stepped-impedance resonator.
The ABCD matrix associated with the complete resonator structure is therefore:
\begin{equation}
	\left[ \begin{array}{cc}A&B\\C&D\end{array} \right]_{res} = \left[ \begin{array}{cc} 1 & 0\\ 1/Z_{tot} & 1 \end{array} \right].
	\label{eq:ABCD_Y}
\end{equation}
The individual resonators are connected to the feedline via a T-junction at their respective electrical delays. The following expression,
\begin{equation}
	\left[ \begin{array}{cc} A&B \\ C&D \end{array} \right]_{T-junc} = \left[ \begin{array}{cc} A&B \\ C&D \end{array} \right]_{TL} \left[ \begin{array}{cc} A&B \\ C&D \end{array} \right]_{res} \left[ \begin{array}{cc} A&B \\ C&D \end{array} \right]_{TL},
	\label{eq:ABCD_Tjunc}
\end{equation}
allows the resonators to be cascaded with the feedline sections, whose ABCD matrices are expressed through Eq.~\eqref{eq:ABCD_TL}. The resulting two-port circuit is used to compute the frequency response of the chip.
The wire bonds, transition board, connectors, and coaxial cables between the device reference and the instrument termination planes were added in a similar fashion to investigate the baseline response. For simplicity we do not explicitly treat the parasitic reactance at the transmission line junctions; however, this detail can readily be incorporated in the formulation and can be of importance at higher frequencies. The complex phase velocity and kinetic inductance fraction of the transmission line configuration employed can be extracted from the response of this composite ABCD-matrix for the system.

\bibliography{Bibliography}

\end{document}